\documentclass[a4paper,fleqn,usenatbib]{mnras}

\usepackage[T1]{fontenc}
\usepackage{ae,aecompl}
\usepackage{graphicx}	
\usepackage{amsmath}	
\usepackage{amssymb}	

\usepackage{natbib}
\usepackage{graphicx}
\title[SWIFT view of V404 Cyg]{SWIFT view of the 2015 outburst of GS 2023$+$338 (V404 Cyg): complex evolution of spectral and temporal characteristics}

\author[Radhika D. et al.]{
Radhika D.,$^{1}$\thanks{E-mail:radhikad\_isac@yahoo.in }
Nandi A.,$^{2}$
Agrawal V. K.$^{2}$
and Mandal S.$^{3}$
\\
$^{1}$Department of Physics, University of Calicut, Malappuram (District), 673635, India\\
$^{2}$Space Astronomy Group, ISRO Satellite Centre, ISITE, Outer Ring Road, Near Karthik Nagar, 
Marathahalli, Bangalore, 560037, India\\
$^{3}$Indian Institute of Space science and Technology, Trivandrum, 695547, India \\
}

\date{Accepted XXX. Received YYY; in original form ZZZ}
\pubyear{2016}

\begin{document}
\label{firstpage}
\pagerange{\pageref{firstpage}--\pageref{lastpage}}
\maketitle

\begin{abstract}
We study the spectral and temporal characteristics of the source GS 2023$+$338 (V404 Cyg) 
during the initial phase of its June 2015 outburst, over the energy range of 0.5 - 150 keV. This is the 
first detailed study of the characteristics of this source based on {\it SWIFT} observations, being 
reported. Based on our analysis, we understand that the source existed in the hard, intermediate and 
soft spectral states. We find that the evolution of the spectral parameters, the hardness intensity 
diagram and the rms-intensity diagram are not similar to those observed for most of the 
outbursting black hole sources. We also observe presence of weak peaked components in the power density 
spectra during the intermediate state of the source. Dramatic changes in the spectral and 
temporal properties are also exhibited before the ejection of a radio jet suggesting it to be associated 
with the coronal mass ejection. It seems that may be due to evacuation of the inner part of the 
Keplerian disc for a short duration, the disc component is not observed after the huge 
radio flare. The absorption features observed in the low energy spectra suggest the presence of wind 
emission and the evolution of the characteristics of the variable Fe line emission during both hard and 
intermediate states, indicate its origin to be probably related to the wind/outflow.

\end{abstract}


\section{Introduction}
\label{intro}

Black hole transients have been observed to depict variability in spectral and temporal 
characteristics during their outbursts. The light curves for the outbursts of these sources are 
observed to be typically of fast rise and exponential decay (FRED) profile (e.g. XTE J1859$+$226 
\citep{Brock2002}, 4U 1543$-$47 \citep{2004ApJ...610..378P}), but a few sources are observed to show 
deviations from this (e.g. GRO J1655$-$40 \citep{1998ApJ...494..753K}, GX 339$-$4 \citep{Belloni2005}). 
The thermal emission from the Keplerian accretion disc around these objects contributes to the 
soft spectra \citep{1973A&A....24..337S}. The high energy spectra are resultant of the 
Comptonization of the seed photons from the disc by the hot electrons in the corona which 
may be static \citep{1995xrbi.nasa..126T} in nature or a dynamical corona \citep{ST95} that satisfactorily 
explains the spectral evolution observed in the outbursting black hole sources. The resultant high energy
spectra, can some times have a cut-off at higher energies corresponding to the recoiling of the electrons. 

The variation in contribution from different spectral components can give rise to spectral 
and temporal variabilities and based on these observed characteristics, black hole 
transients undergo spectral state transitions in their Hardness Intensity Diagram (HID) which 
typically has a `q-shape' for most of the sources 
(\citealt{Belloni2005,2010MNRAS.403...61D,2016A&A...587A..61C} and references therein). 
\citealt{2006ARA&A..44...49R} classified these states into soft (HSS), hard (LHS) and very high states, 
while \citealt{Belloni2005,Motta2012} and references therein further classified the very high state 
into hard-intermediate (HIMS) and soft-intermediate (SIMS) states with more emphasis on the 
timing properties of the sources (eg. types of QPOs, fractional rms variability etc.). 
With reference to these understandings of the combined spectral and temporal properties, 
\citealt{Deb08,Nandi2012,RN2014} have studied the evolution of spectral states in HID for several sources.
The underlying principle of classification of states was purely based on the prescription of two 
component advective flow (\citealt{ST95}) model (see section 4 for details). Recently, this classification
scheme is further corroborated based on the physical basis of accretion dynamics around several black hole
sources (\citealt{2015MNRAS.447.1984D,2015ApJ...803...59D,2015ApJ...807..108I,2016MNRAS.tmp..664M} 
and references therein).
  
In outbursting black hole sources the variation of source intensity with respect to the 
fractional rms variability is observed to show a hysteresis \citep{Munoz11,KB15} which is 
called as the rms-luminosity diagram (RLD). It has also been understood that a steady/compact jet exists 
during the LHS and HIMS (\citealt{FBG04,FHB09} and references therein), while usually a transient 
relativistic jet is observed during the transition from HIMS to SIMS. Jets are not observed during the 
HSS. Hence, the characteristics of the source during the jet ejections help in
understanding of the phenomenon of accretion-ejection or disc-jet coupling. The
spectral and temporal characteristics of most of the outbursting black hole binaries, 
have been understood based on the continuous observations by Rossi X-ray Timing Explorer (RXTE) 
\citep{Belloni2005,2006ARA&A..44...49R,Nandi2012}.

The source GS 2023$+$338 is a black hole transient with a low mass optical companion V404 Cyg. It was 
discovered by the Ginga satellite during its outburst in 1989 \citep{1989IAUC.4782....1M}, and is 
located at an R.A. of 306.01$^{\circ}$ and DEC of 33.86$^{\circ}$. The dynamical mass and distance of the 
source are estimated to be of ${\sim}$ 12M$_{\odot}$ \citep{1994MNRAS.271L..10S}, and 
2.39 ${\pm}$ 0.14 kpc \citep{2009ApJ...706L.230M}, respectively. 

The source has been observed to show variabilities in its light curve, by undergoing multiple flaring 
activity. The light curve of the source has been modeled as a triangular profile during its 1989 
outburst \citep{1989Natur.342..518K}. The X-ray spectral and temporal characteristics of this 
source have been studied based on GINGA \citep{1989Natur.342..518K} and ROENTGEN observations 
\citep{1992A&A...266..283I,1997A&A...321..776O}, highlighting the changes in spectral index, variable 
low energy absorption and slight variations in the power spectrum with respect to typical black hole 
sources. The radio and X-ray correlation for the source during its 1989 outburst and quiescence phase 
has been performed by \citealt{2008MNRAS.389.1697C}, 
and they observed the linear correlation to be consistent during the hard state of the source in its 
1989 outburst and also later at the quiescence. 

A recent outburst of V404 Cyg was detected by SWIFT-Burst Alert Telescope (BAT) on 15th June 
2015 (i.e. MJD 57188), with follow-up observations by SWIFT-X-ray Telescope (XRT) 
\citep{Barthelmy2015}, Monitor of All-sky X-ray Image (MAXI) \citep{2015ATel.7646....1N}, 
INTErnational Gamma-Ray Astrophysics Laboratory (INTEGRAL) \citep{2015ATel.7647....1K} and 
FERMI Gamma-ray Space Telescope (hereafter FERMI) \citep{2015GCN..17932...1Y}. Multi-wavelength 
observations were performed in Optical \citep{2015ATel.7650....1G} and Radio 
\citep{2015ATel.7658....1M,2015ATel.7716....1T}. In contrary to most of the black hole 
transients having typically one X-ray flare, for the source V404 Cyg multiple X-ray flares are 
observed during the 2015 outburst similar to its 1989 outburst \citep{2015ATel.7662....1F}. 
The source achieves a peak X-ray flux approximately 10 days after the beginning of the outburst and 
later decays to the quiescence phase \citep{2015ATel.7763....1S}. Detailed study of 
the spectral characteristics observed by INTEGRAL has been discussed by 
\citealt{2015A&A...581L...9R,2015ApJ...813L..21N,2015ApJ...813L..22R}. They classified the spectral 
states into hard `on-flare' and hard `off-flare' states (i.e. hard spectra when a flare occurs 
and does not, respectively), and also obtained the seed photon temperature required for 
Comptonization to be $>$7 keV. Analysis of the FERMI-Gamma ray Burst Monitor (GBM) 
observations by \citealt{2016arXiv160100911J} also suggested the source spectrum to be hard during the 
initial days of the outburst. Although the nature of the hard X-ray spectrum was understood by these 
studies, a detailed study of the contribution from the soft/thermal disc emission was not 
reported. Based on the INTEGRAL observations, the X-ray variability observed was attributed to the 
accreting source and also to the variable absorption by matter in the line of sight which was 
seen during the 1989 outburst too \citep{1992A&A...266..283I}. The Chandra-HETG observations have 
revealed the detection of emission lines in the spectrum of the source during the 2015 outburst, 
indicating strong disc wind emission \citep{2015ApJ...813L..37K}. 

Previous studies on this source based on the 1989 outburst, found that the source exhibited only a 
hard spectrum without any signature of the Keplerian disc. The timing analysis implied that the 
power density spectrum had flat-top noise similar to that observed during the hard states of black 
hole binaries. The spectral and temporal characteristics of the source were then 
correlated to the black hole source Cyg X$-$1 \citep{1992A&A...266..283I,1997A&A...321..776O}. The 
INTEGRAL (\citealt{2015ApJ...813L..22R} and references therein) and FERMI \citep{2016arXiv160100911J} 
observations of the 2015 outburst could classify the source into a hard state only. 

In this paper, we consider SWIFT XRT and BAT observations of GS 2023$+$338 (V404 Cyg) during the initial 
phase of its 2015 outburst, so as to understand the spectral and temporal characteristics of 
the source when multiple X-ray and radio flares are observed. We consider the energy range of 
0.5 - 150 keV, so as to look into the possible contributions of both the soft and hard emissions from 
the source. This paper for the first time gives a detailed analysis of the SWIFT XRT and BAT 
observations for the 2015 outburst of V404 Cyg. Although the SWIFT observations are not 
continuous unlike RXTE observations of outbursting sources\footnote{V404 Cyg has not been 
observed by RXTE, but in this paper we discuss the spectral and temporal characteristics of 
the source based on the general understanding about black hole binaries that we have learned 
using RXTE observations.}, these are good enough to give an overall understanding of the evolution 
of the source characteristics, spectral states and HID. The previous outburst of this source has not 
exhibited any spectral state transition. We attempt to understand if any state transition takes place 
during the 2015 outburst. This is investigated by studying the variations of flux from the soft and 
hard components, hardness ratio and the fractional rms variability observed by both XRT and 
BAT. We attempt to understand the evolution of the Fe-line emission, and presence of any absorption 
features during this outburst. We also explore the possible correlation between the X-ray 
characteristics and reported radio emissions. 

A summary of the procedures followed for data reduction and analysis has been given in section \ref{obs}. 
The results obtained from the spectral and temporal analysis are presented in section \ref{res} 
and the same are discussed in section \ref{dis}. 

\section{Observations, data reduction and analysis}
\label{obs}
We have analyzed the public archival data of the SWIFT satellite, available through the HEASARC database. 
Data are obtained for 23 observations beginning from MJD 55188 (15th June 2015) and up-to 
MJD 55201.99 (28th June 2015) when on-wards the dust scattering rings are observed 
\citep{2015ATel.7736....1B, 2016MNRAS.455.4426V}. Amongst these, simultaneous XRT and BAT observations 
exist for eight observations (i.e. MJD 57191.03, 57191.54, 57194.60, 57196.34, 57198.01, 57199.52, 
57199.99 and 57200.92), four (i.e. MJD 57188.77, 57193.55, 57197.20 and 57198.15) have 
only BAT observations while the rest have only XRT. An observation log has been tabulated as 
part of the appendix. In this paper, we refer MJD 57188 as day 0, and all the other observations 
follow accordingly. The standard ftools provided by HEASOFT v 6.17 are used for the purpose of data 
reduction and analysis.   

\subsection{XRT data reduction}
The SWIFT X-ray telescope (XRT) is a Wolter-type 1 telescope with a CCD at its focus and covers an 
energy range of 0.2 - 10 keV \citep{2005SSRv..120..165B}. Amongst the different read-out modes of XRT, 
for the source V404 Cyg the observations are performed in the windowed-timing mode.

We obtain the cleaned XRT event products through the \emph{xrtpipeline}. Using XSELECT v 2.4, events are 
generated corresponding to grades of \emph{0-2} which consist of the valid X-ray events. For events 
without pile-up, a circle of radius 30 pixels is chosen for the source 
region. Since the source has been bright at a few intervals, the pile-up corrections are performed for 
these particular observations by selecting events from grades of only \emph{0}. In this case, an annular 
region is chosen around the source centre. The outer radius of the annular is fixed at 30 pixels and 
the inner radius is selected by varying the number of pixels, based on the ratio of counts in grade 0 
events to grade 0-2 events. An annular 
background region is considered far away from the source, referring to the standard methods for 
windowed-timing mode data. For the observations where the source has 
been contaminated by dust scattering, the background region is selected in such a way that it will 
subtract the contribution from the dust (A. Beardmore, 2015 private communication). 

Since the extraction regions are not always similar, a scaling factor has been applied to the source and 
background spectral files, by editing the BACKSCAL keyword 
\footnote{http://www.swift.ac.uk/analysis/xrt/backscal.php}. If the source region is circular with 
radius r$_s$, then a scaling factor of 2$\times$r$_s$ is applied to the source spectra. The annular 
region chosen for the background has inner radius r$_1$ and outer radius of r$_2$, so the scaling 
factor applied is r$_2$ $-$ r$_1$. The pixels at the end-of-window will be considered as bad by the 
ground software processing, thereby converting the image from 2-dimensional to 1-dimensional. 
The BACKSCAL keyword takes into account these and corrects the spectral information. 

Uncertainty in position of the source has been taken care of, by applying the position dependent 
\emph{rmfs} for the respective grades (see also \citealt{2015ApJ...807..108I}). Those observations 
where the source is not piled-up, the rmfs 
used do not have these corrections \footnote{http://www.swift.ac.uk/analysis/xrt/rmfs.php}. The ftool 
\emph{xrtmkarf} has been used along-with the exposure map, to generate the arf files for each 
observation. The source spectral data are re-binned to 25 counts per bin, by means of \emph{grppha}. 

Timing analysis of XRT data is performed by generating source and background light curves for the 
minimum time bin resolution of 1.8 ms, for the selected source and background regions. From these, we 
generate background subtracted light curves and use for further analysis. 

\subsection{BAT data reduction}

The Burst Alert Telescope (BAT), has a large area solid state (i.e. CdZnTe) detector with energy range 
coverage of 15 - 150 keV and a large field of view \citep{2013ApJS..209...14K}. The instrument is a very 
good monitor for the bursting sources (say GRBs, transients etc.), and also has good capabilities to 
study the spectral and temporal variabilities from a source. 

The standard procedures for data reduction mentioned in the data analysis threads 
\footnote{http://www.swift.ac.uk/analysis/bat/}, have been employed to generate BAT spectra and 
light curves. We consider the event data for the analysis, referring to the analysis procedures 
provided in the BAT threads. Since the gain correction has already been applied in the BAT event file, 
we did not repeat the energy calibration 
\footnote{http://swift.gsfc.nasa.gov/analysis/threads/bateconvertthread.html}. Detector plane images 
(dpi) are generated for the energy range of 15 - 150 keV using \emph{batbinevt}. The
 \emph{batdetmask} is used to obtain the appropriate detector quality map with the help of the 
detector enable/disable map. Making use of these, the noisy detectors are found 
with the help of the ftool \emph{bathotpix}, and a quality map is obtained as a result. Appropriate 
mask weighting is applied to the event data using \emph{batmaskwtevt}. We generate energy spectra 
for the range of 15 - 150 keV by incorporating the quality map. Systematic errors are applied 
using \emph{batphasyserr} to account for the residuals in 
the response, the ray-tracing keywords are corrected by \emph{batupdatephakw}. Response matrix 
corresponding to the spectral file has been generated using \emph{batdrmgen}. 

BAT light curves are generated for a time bin of 0.01 sec over the energy range of 15 - 150 keV, with 
the help of the ftool \emph{batbinevt}. 

\subsection{Analysis}

Spectral analysis package of XSpec v 12.9 \citep{Arn96} is used to perform simultaneous spectral analysis 
of the XRT and BAT data over an energy range of 0.5 - 150 keV, which we find to be the optimum energy 
range having statistically significant photon counts required for spectral fitting. Spectral modeling 
has been performed with the help of the \emph{diskbb} \citep{1984PASJ...36..741M,Maki86}, \emph{compTT} 
\citep{1994ApJ...434..570T} or \emph{pexrav} \citep{1995MNRAS.273..837M}, \emph{gauss}, \emph{pcfabs} and 
\emph{gabs} models. 

The \emph{diskbb} is used to model the low-energy photon emission from different radii of the 
Keplerian accretion disc. This model provides an estimate of the temperature at each radius of 
the disc (hence, a multi-coloured disc with blackbody spectra) and the contribution 
to the flux emitted in soft energy. The high energy spectrum will be a powerlaw which 
is occurring due to Comptonization of the soft photons by the corona. The \emph{compTT} model describes 
the Comptonization process of relativistic thermal plasma. The \emph{pexrav} considers a 
powerlaw spectrum with exponential cut-off, which gets reflected from a neutral material. We study the 
evolution of the photon index of the spectra, the reflection factor and the cut-off energy in order 
to understand the characteristics of the hard emission. Detailed information about the different models 
used in this paper can be obtained from the corresponding references and also in the XSpec manual. 

The \emph{gauss} model is used to study the fluorescent Fe-line emission signature in the spectrum, and 
the absorption features of the spectra are modeled by \emph{pcfabs} and/or \emph{gabs} 
(see section \ref{res} and \S \ref{dis}). The interstellar absorption has been considered 
using the \emph{wabs} model \citep{2000ApJ...542..914W}, and the hydrogen column density 
(n$_H$ factor) is found to be varying between 0.6$\times$10$^{22}$ atoms cm$^{-2}$ and 
1.2$\times$10$^{22}$ atoms cm$^{-2}$, during the outburst. 

The \emph{cflux} model is used to obtain the source X-ray flux (i.e. unabsorbed flux) in the 
energy ranges 0.5 - 10 keV and 15 - 150 keV. The contribution of disc 
flux is estimated in 0.5 - 10 keV and that of the hard/powerlaw flux in 0.5 - 10 keV and/or 15 - 150 keV. 
We estimate the hardness ratio in XRT and BAT, by calculating the ratio 
of fluxes in 4 - 10 keV and 0.5 - 4 keV,  
50 - 150 keV and 15 - 50 keV, respectively. Error bars for the 
different spectral parameters are obtained at 90\% confidence interval, using the command \emph{err}. 

Timing analysis has been performed using the package XRONOS v 5.22. The frequency range 
considered is 0.068 Hz to 277.8 Hz for the XRT data of 1.8 ms time resolution and interval 
length of 8192 bins, and 0.012 Hz to 50 Hz for BAT data of time resolution 0.01 sec and interval length 
of 8192 bins. 
In order to search for the presence of QPOs, power density spectra (PDS) are generated 
for the same frequency range, using the ftool \emph{powspec} v 1.0 by performing Fourier transform on 
the background subtracted lightcurves. The fractional rms variability is estimated by calculating the 
square root of the integrated power in the XRT lightcurve for the frequency range of 0.1 - 20 Hz, 
thereby neglecting the noise dominant frequency range. We calculate the error in rms variability 
at 90\% confidence interval, by means of error propagation. Since BAT PDS has broad-band noise with 
less power, we did not estimate the rms variability for the same. We generate the XRT PDS with 
interval length of 65536 bins, resulting the frequency range to begin from 0.008 Hz, so as to search 
for mHz QPOs.

The Radio data have been obtained by Radio Astronomical Telescope of the Academy of 
Sciences (RATAN)-600, and we refer to the radio light curve provided by them \citep{2015ATel.7716....1T}. 
Radio flaring has also been reported by \citealt{2015ATel.7658....1M,2015ATel.7714....1M} based 
on Arcminute Microkelvin Imager - Large Array (AMI-LA) observations at 16 GHz, 
\citealt{2015ATel.7701....1T} using Waseda University Nasu radio telescope at 1.4 GHz and 
\citealt{2015ATel.7708....1T} using Very Large Array (VLA) at different frequencies. The possible 
correlation between X-ray characteristics and radio flares are explored in section 3.4.

\section{Results}
\label{res}

We have performed a detailed spectral and temporal analysis of the source V404 Cyg, and the 
results of the same are presented here.

\subsection{Spectral evolution}
\begin{figure}
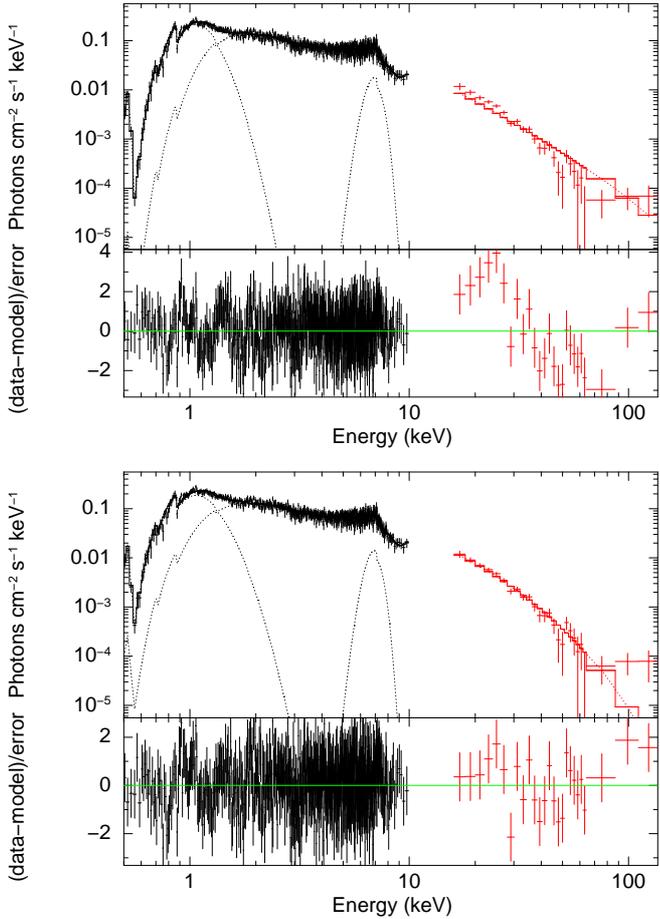

\begin{minipage}{9cm}
\includegraphics[height=9cm,angle=-90]{fig1a.ps}
\end{minipage}
\vskip 0.1in
\begin{minipage}{8cm}
\includegraphics[height=9cm,angle=-90]{fig1b.ps}
\end{minipage}
\caption{Spectral fit for simultaneous XRT and BAT observation on day 10.01 (MJD 57198.01) over the 
energy range of 0.5 - 150 keV, using the model \emph{wabs*(diskbb+gauss+compTT)} in top panel and 
\emph{wabs*(diskbb+gauss+pexrav)} in bottom panel. These fits indicate the presence of disc and 
Fe line components. The solid lines represent the fit to the data, while the dotted lines denote the 
model components. The BAT spectra has been re-binned by 10 counts/bin above 60 keV for better 
representation.}
\label{fig1}
\end{figure}

We carry out the spectral analysis, by fitting the energy spectrum in 0.5 - 150 keV, using several models. 
The spectral fit with all the spectral components for one of the simultaneous observations 
(i.e. day 10.01; MJD 57198.01), has been shown in Figure \ref{fig1}. We find that the spectral fit 
using \emph{diskbb}, \emph{gauss} and \emph{compTT} (see top panel of Figure \ref{fig1}) results 
in $\chi^2$/dof of 791/689, and gives large residuals above 15 keV indicating the presence of a 
reflection component. We also could not constrain the \emph{compTT} parameters. Hence, we replace 
\emph{compTT} with the \emph{pexrav} model which takes into account the reflection component and any 
high energy cut-off observed. This improves the fit and results in $\chi^2$/dof of 699/687. 
The \emph{diskbb} model is used for those observations where we find presence of a disc 
component in the low energy spectrum. We also find evidence of prominent Fe line emission since day 3 
(i.e. MJD 57191) during seven out of the entire twenty observations, and it is modeled using \emph{gauss}. 
The inclination angle of the disc has been estimated as 56${\pm 4}$$^{\circ}$ 
by \citealt{1994MNRAS.271L..10S}, and we refer to the same. The final model used is 
\emph{wabs*(diskbb+gauss+pexrav)} as shown in the bottom panel of Figure \ref{fig1}.  

\begin{figure}
\includegraphics[width=9cm]{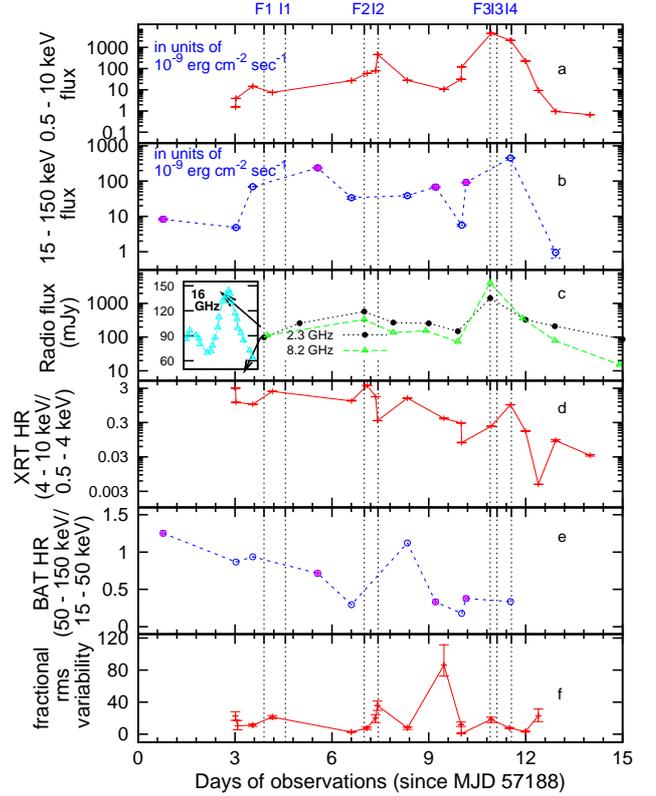}
\caption{Variations of integrated XRT flux (a), BAT flux (b), Radio flux (c), Hardness ratios for XRT (d), 
Hardness ratios for BAT (e) and fractional rms variability in \% (f) for XRT, are shown in different 
panels of this plot. 
Here XRT observations are shown in red cross, BAT observations by magenta box and simultaneous 
XRT+BAT observations by blue open circular points whereas radio flux are shown by black dots and 
green diamond points. Error bars for the different data points are also marked, and some of the 
observations have error bars smaller than the symbol size. The Radio flare 
detected by AMI-LA at 16 GHz during day 4 (i.e. MJD 57192) has been shown in the inset of the panel c, 
where the observations (cyan diamond points) are quoted in hours scale. F1, F2 and F3 represent the 
radio flares observed and I1, I2, I3 and I4 denote the X-ray flares detected by INTEGRAL. In the X-axis 
of the figure, day 0 corresponds to MJD 57188.}
\label{fig2}
\end{figure}

\begin{figure}
\includegraphics[width=\hsize]{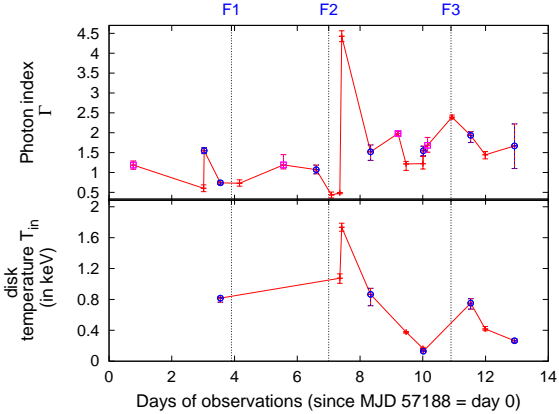}
\caption{Evolution of the spectral parameters along-with their error bars, during the outburst for XRT 
(red cross points) and BAT (magenta box type points) observations. Simultaneous observations have been 
marked in blue open circular points. In the X-axis of the figure, day 0 corresponds to MJD 57188.}
\label{fig3}
\end{figure}

Figure \ref{fig2} represents the variation of the X-ray flux in 0.5 - 10 keV and 15 - 150 keV (panels a 
and b respectively), the radio flux (panel c) observed by RATAN-600 \citep{2015ATel.7716....1T}, 
variation of hardness ratio (panels d and e) and fractional rms variability estimated from XRT 
observations in the frequency range of 0.1 - 20 Hz (panel f). The data points corresponding to the XRT 
observations are shown in red cross type points, whereas BAT in magenta box type points 
and the simultaneous XRT-BAT observations are indicated in blue open circular points. Error 
bars for all the observed/estimated parameters are also included. Some of the observations 
have error bars smaller than the symbol size. The different radio flares reported have been 
marked as F1, F2, F3. Around sixteen flares in X-rays have been detected by INTEGRAL 
(see \citealt{2015A&A...581L...9R} and INTEGRAL public data products), with flux $>$ 6 Crab 
(i.e. $>$ 990 counts/sec in 20 - 40 keV). Amongst these flares, we have 
represented in Figure \ref{fig2}, those flares which are closer to the X-ray flares we obtained from 
SWIFT analysis, and they have been indicated as I1, I2, I3 and I4. Variations of the different 
spectral parameters are shown in Figure \ref{fig3}, with representation of data points similar to 
Figure \ref{fig2}. In Table \ref{all} we present model fitted spectral parameters from all the observations.

In the beginning of the outburst for a few observations (i.e. days 0.77, 3.01 and 
3.03), we find that the spectra can be fitted well by a powerlaw. For uniformity, we have modeled 
the spectra with \emph{pexrav}, which indicates 
lack of cut-off and reflection components for these observations. The observations after 3 days of 
the outburst (i.e. since MJD 57191.54), require a \emph{diskbb} model along-with the \emph{pexrav} 
model to fit the data, implying the presence of a soft disc and reflection component together-with the signature of high energy cut-off for simultaneous XRT and BAT observations.

We observe a similar spectral nature with the presence of both disc and hard 
components for all the observations considered, except for a few. During days 4.16, 6.40, 
7.08 and 10.93, the XRT spectrum fits well 
with the \emph{pexrav} without any \emph{diskbb} component. We notice that this intermittent absence 
of disc component takes place just after a radio flare (see section 4) and 
makes this source very different from most of the other black hole transients. 

We find that since day 3.54 (see also Table \ref{all}), the spectral fits for all the observations 
(except days 12 and 13) in the low energy range of 0.5 - 10 keV, require a model corresponding to 
absorption along-with the \emph{diskbb} or/and \emph{pexrav}, and \emph{wabs} models. This absorption 
feature has been taken care using the \emph{pcfabs} model, with the covering fraction varying between 
0.5 and 0.95, to fit the spectral data. As an example for day 10.00, we find that the inclusion 
of \emph{pcfabs} improves ${\chi}^{2}$/dof from 918/667 to 725/665 for the low energy XRT spectral 
fit, suggesting this absorption characteristic to be statistically significant. In addition to this 
spectral absorption, we observe absorption lines of peak energy $\sim$ 0.67 keV with width 
0.1 keV on days 6.60 and 8.34, while $\sim$ 9 keV absorption line of width $\sim$ 0.78 keV 
is observed on the 10th day of observations. We model both of these lines\footnote{We observe 
an indication for these absorption lines, but due to lower photon statistics (because of less effective 
area of XRT) an accurate modeling of the same is not possible.} using \emph{gabs}.  

Although we did observe the contribution from Keplerian disc for a few observations, Figure 
\ref{fig3} and Table \ref{all} indicate that the variation of disc parameters is random. 
The disc temperature is observed to vary between 0.16${\pm 0.01}$ keV to 1.73${\pm 0.05}$ keV. We note that the disc temperature is around 0.1 keV to 0.3 keV for few observations (although XRT
spectra are modelled from 0.5 keV onwards), but these observations also have large contribution of the 
soft flux. The spectral fits show a significant presence of a disc component during these 
observations, say with the $\chi^2$/dof improving from 1039/667 to 725/665 on day 10.00. The lower value of
temperature could be an estimate of the physical disc temperature due to the extrapolation of 
the fit to the spectra. Also, the variation in the uncertainty of disc temperature is lesser 
than the disc temperature itself. 

We find that the photon index has a random variation between 0.43${\pm 0.07}$ to 4.43$^{0.27}_{-0.14}$ 
(Figure \ref{fig3} and Table \ref{all}), unlike the typical black hole sources where it exhibits 
continuous increase (e.g. GX 339$-$4 \citep{Belloni2005}, GRO J1655$-$40, XTE J1550$-$564, 
XTE J1650$-$500 \citep{2009ApJ...699..453S}, XTE J1859$+$226 \citep{RN2014}, 
IGR J17091$-$3624 \citep{2015ApJ...807..108I}) during the initial phase of the outburst.

The XRT flux (0.5 - 10 keV) reaches to a maximum on day 10.93 (Figure \ref{fig2}), with an observed 
value of 4560.0${\pm 23.3}$ ${\times}$10$^{-9}$ erg cm$^{-2}$ sec$^{-1}$ corresponding to a 
luminosity of 2.96${\pm 0.4}$ ${\times}$10$^{39}$ erg sec$^{-1}$ for the source distance of 2.39 kpc 
(see also section \ref{intro}) estimated by \citealt{2009ApJ...706L.230M}. 
On day 11.5, the 0.5 - 10 keV X-ray flux is 2130.7${\pm6.5}$ ${\times}$10$^{-9}$ erg cm$^{-2}$ sec$^{-1}$ 
and the 15 - 150 keV BAT flux achieves a peak (Figure \ref{fig2}) of 
453.1${\pm 5.1}$ ${\times}$10$^{-9}$ erg cm$^{-2}$ sec$^{-1}$ (see also \citealt{2015ATel.7755....1S}). 
The corresponding peak BAT luminosity is 0.29${\pm 0.04}$ ${\times}$10$^{39}$ erg sec$^{-1}$. The total 
X-ray flux in 0.5 - 150 keV during day 11.5, is found to be 
2583.8${\pm 8.24}$ ${\times}$10$^{-9}$ erg cm$^{-2}$ sec$^{-1}$, which is equivalent to a luminosity 
of 1.68${\pm 0.08}$ ${\times}$10$^{39}$ erg sec$^{-1}$. \citealt{2015ATel.7755....1S} reported a maximum 
luminosity of the order of 10$^{39}$ erg sec$^{-1}$ for a 12M$_{\odot}$ central source 
(see section \ref{dis}).

\begin{table*}
\caption{Best fitted model parameters representing the different spectral properties. Here `Day' corresponds 
to days since MJD 57188 (day 0), T$_{in}$ is disc temperature, $\Gamma$ is photon index, fE is 
fold energy in keV and rel$_{refl}$ is the reflection factor and EW is equivalent width of the 
Fe line. Both the disc flux and hard flux, are quoted in units of 
10$^{-9}$ erg cm$^{-2}$ sec$^{-1}$. The spectral parameters 
mentioned are same as that represented in Figure \ref{fig3}. The contribution of disc flux is 
estimated in 0.5 - 10 keV energy range (i.e. range A), while that of the powerlaw flux is estimated 
in 0.5 - 10 keV (range A) for the observations with only XRT data, and in 15 - 150 keV (range B) for 
the rest. The parameters of fE and rel$_{refl}$ are quoted only for the simultaneous observations where 
the cut-off and reflection features are statistically significant.}
\label{all}

\centering
\begin{tabular}{lclcccccccc}
\hline\\
Day & T$_{in}$ & disc flux & $\Gamma$ & \multicolumn{2}{c}{powerlaw$^{\ast}$ flux} & line energy & EW & fE & rel$_{refl}$ & ${\chi}^{2}$/dof $^{\bullet}$   \\
\cline{5-6} \\
    & (keV)      & (range A)  &          & (range A) & (range B)   & (keV)     & (keV)& (keV) & \\         
\hline\\
0.77 & - & - & 1.18${\pm 0.1}$ &-& 8.78${\pm 0.41}$ & - & - & - &-&47.43/36\\\\
3.01 & - & - & 0.60$^{0.08}_{-0.09}$ & 1.43${\pm 0.03}$ &-& 6.60${\pm 0.08}$ & 0.77$^{+0.22}_{-0.19}$ & -&-& 253/282   \\\\
$^{\dagger}$3.03 & - & - & 1.55$^{0.08}_{-0.09}$&-&4.83$^{0.04}_{-0.05}$&-&-&-&-&286/410\\\\ 
$^{\dagger}$3.54$^{\star}$ &0.82$^{0.01}_{-0.05}$& 5.15${\pm 0.04}$ &0.74$^{0.09}_{-0.04}$& - & 69.44${\pm 1.0}$ & - & - & 42.43$^{+0.05}_{-0.03}$ & -&887/1111\\\\ 
4.16$^{\star}$ & - & - & 0.73$^{0.09}_{-0.07}$ & 7.49${\pm 0.07}$ & - & 6.26$^{0.1}_{-0.06}$ & 1.15$^{+0.22}_{-0.19}$ & - & -& 933/753  \\\\
5.56$^{\star}$ & - & - & 1.19$^{0.2}_{-0.1}$ &-& 241.28${\pm 2.7}$ & - & - & - &-&43/71 \\\\
$^{\dagger}$6.60$^{\star}$ & - & - & 1.07${\pm 0.08}$ &- & 33.81${\pm 3.1}$ & 6.84${\pm 0.08}$ & 0.30${\pm 0.07}$ & 18.81$^{+4.2}_{-3.3}$ & 0.67${\pm 0.23}$ & 912/1096\\\\
7.08$^{\star}$ & - & - & 0.43${\pm 0.07}$ & 55.13${\pm 0.38}$&- & 6.73${\pm 0.04}$ & 0.47$^{+).07}_{-0.06}$ & - & -& 931/820  \\\\
7.35$^{\star}$ & 1.07$^{0.6}_{-0.7}$ & 16.68${\pm 0.01}$ & 0.48$^{0.03}_{-0.02}$& 62.81${\pm 0.27}$&- &-&-&-&-& 1405/914\\\\
7.41$^{\star}$ & 1.73${\pm 0.05}$&304.92${\pm 1.51}$&4.43$^{0.27}_{-0.14}$& 155.19${\pm 3.1}$&- &-&- &- &-&973/710\\\\
$^{\dagger}$8.34$^{\star}$ & 0.86$^{0.15}_{-0.08}$ & 4.95${\pm 0.05}$ & 1.52$^{0.18}_{-0.21}$& - & 36.53${\pm 3.0}$ & 6.59${\pm 0.05}$ & 0.71$^{+0.10}_{-0.07}$ & 100.90$^{+14.8}_{-12.7}$ & 10.01${\pm 0.87}$ &832/952 \\\\
9.21$^{\star}$ & - & - & 1.98${\pm 0.05}$ &-& 68.15${\pm 3.7}$ &-&-&-&-&32/36 \\\\
9.47$^{\star}$ & 0.38${\pm 0.01}$ & 5.22${\pm 0.19}$ & 1.22$^{0.22}_{-0.16}$ & 5.35${\pm 0.13}$&- & 6.43${\pm 0.04}$ & 0.23$^{+0.4}_{-0.05}$ &-& -&311/347  \\\\
10.00$^{\star}$ & 0.16${\pm 0.01}$$^{!}$ & 18.54${\pm 0.49}$ & 1.22$^{0.3}_{-0.1}$& 12.73${\pm 0.12}$ &-&-&-&-&-&725/665 \\\\
$^{\dagger}$10.01$^{\star}$ & 0.13${\pm 0.01}$$^{!}$ & 105.91${\pm 3.3}$ & 1.54$^{0.2}_{-0.1}$&-&5.64${\pm 0.28}$ &6.9$^{\$}$&0.29&16.95$^{+5.01}_{-2.32}$ &0.97$^{0.15}_{-0.16}$& 699/687 \\\\
10.15$^{\star}$ & - & - & 1.68$^{0.19}_{-0.17}$ &-& 94.05${\pm 1.2}$ &-&-&-&-&20/37 \\\\  
10.93$^{\star}$ & - & - & 2.39${\pm 0.05}$&4560.0${\pm 23.2}$&- & -&-& -&-&1201/756 \\\\
$^{\dagger}$11.5$^{\star}$ & 0.75$^{0.05}_{-0.07}$& 300.47${\pm 3.57}$ &1.93$^{0.27}_{-0.18}$& 2130.6$^{6.44}_{-6.47}$ &453.11$^{5.2}_{-5.0}$ &-&-&60.79$^{+11.9}_{-16.8}$& 1.39${\pm 0.04}$ &899/1273 \\\\   
11.9$^{\star}$ & 0.42$^{0.03}_{-0.02}$& 160.11${\pm 1.1}$  &1.44${\pm 0.1}$&66.28${\pm 0.76}$&-&-&-&-&-&664/617\\\\
12.39 & 0.49${\pm 0.06}$ & 0.366${\pm 0.09}$ & 4.52${\pm 0.12}$& 8.97$^{0.15}_{-0.59}$& - & - & - & - & - & 397.56/316\\\\
$^{\dagger}$12.92&0.26$^{0.03}_{-0.02}$$^{!}$ & 0.78${\pm 0.03}$ &1.66$^{1.0}_{-0.6}$&-&0.91$^{0.09}_{-0.03}$&-&-&-&-&306/337\\\\
13.99 & 0.25${\pm 0.03}$ &0.37${\pm 0.05}$ & 3.04${\pm 0.4}$ & 0.29${\pm 0.02}$ & - & - & - & - & - & 185.71/177 \\\\
\hline
\end{tabular}
$^{\ast}$ - The \emph{pexrav} model is used for the spectral fit to obtain the contribution from the 
powerlaw/hard flux. \\
$^{\dagger}$ - observations with simultaneous XRT and BAT data.\\
$^{\$}$ - line energy is kept fixed, so as to constrain the parameters.\\
$^{\star}$ - The \emph{pcfabs} model is included while modeling the spectra.\\
$^{\bullet}$ - observations with residuals at Si and Au edges of XRT show ${\chi}^{2}/dof$ $>$ 1.2\\
$^{!}$ - Significant presence of disc component, with increased contribution of soft flux
\end{table*}

\subsection{Temporal evolution}

During the initial days, the XRT power density spectrum shows only broad-band noise and has higher 
fractional rms variability. For days 3.54 and 7.08, we find that the PDS has a powerlaw noise, 
and the fractional rms variability is around 7${\pm 2}$ \% to 11$^{1}_{-2}$\%. The rest of the PDS has 
a broad-band noise. The XRT rms achieves a minimum of 0.79$^{0.5}_{-0.3}$\% on day 10.01 (see panel f of 
Figure \ref{fig2}). A powerlaw shape of the XRT-PDS is observed during day 11 also, but with the 
rms decreased to $\approx$7.6$^{0.4}_{-0.5}$\%.

\begin{figure*}
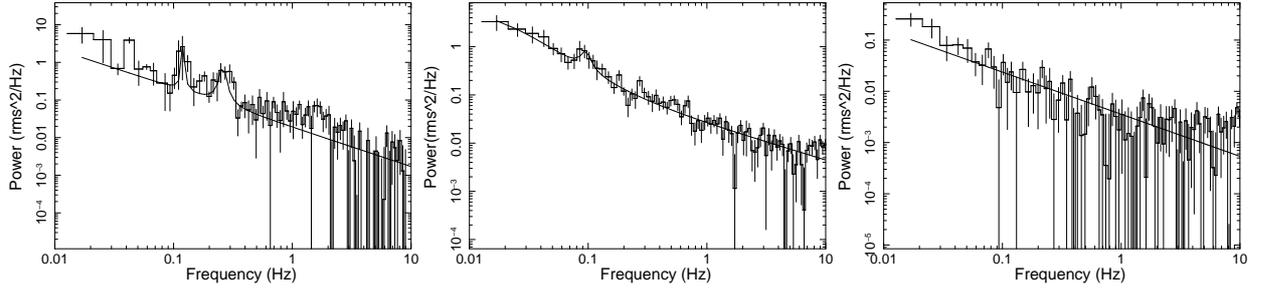

\begin{minipage}{5.2cm}
\includegraphics[height=6.2cm,angle=-90]{fig4a.ps}
\end{minipage}
\hskip 0.06in
\begin{minipage}{5.2cm}
\includegraphics[height=6.2cm,angle=-90]{fig4b.ps}
\end{minipage}
\hskip 0.06in
\begin{minipage}{5cm}
\includegraphics[height=6.2cm,angle=-90]{fig4c.ps}
\end{minipage}
\caption{Power density spectra for the observations on days 3.03, 4.15 and 6.60 respectively. These PDS 
show that there is no clear detection of QPOs but have only weak peaked components. The solid line 
shows the complete fit to the PDS.}
\label{fig4}
\end{figure*}

The INTEGRAL observations performed on days 0 and 2 have reported presence of 
significant low frequency QPOs in their 
PDS \citep{2015ATel.7726....1P} at 0.124 Hz and 0.251 Hz. Low frequency QPOs have also been observed 
by FERMI-GBM \citep{2015ATel.7715....1J,2016arXiv160100911J} during days 4 and 5. We did not 
find any signature of significant ($>$3$\sigma$) QPOs in the PDS of both XRT and BAT, for the 
observations considered. Although, \citealt{2015ATel.7665....1M} reported about the detection of a 
3.1$\sigma$ significant QPO of 1.7 Hz on day 3.03, we do not find any clear detection of this 
QPO during this observation but weak features around 0.12 Hz and 0.25 Hz are evident (see the 1st panel 
of Figure \ref{fig4}). We find that during this period and also when INTEGRAL/FERMI detected QPOs, the 
PDS generated for the minimum time resolution of the SWIFT-BAT observations also do not show significant 
detection of QPOs but only a broad-band noise. FERMI has detected mHz QPOs during days 4 and 5 
\citep{2016arXiv160100911J}. The power spectrum generated for day 4.15, shows a weak 
feature/peaked component at 0.09 Hz (2nd panel of Figure \ref{fig4}). The following XRT observation on 
day 6.60 does not indicate any signatures of peaked components or QPOs but has only broad-band noise 
(3rd panel in Figure \ref{fig4}).

\subsection{Hardness-intensity and rms-intensity diagrams}

\begin{figure}
\begin{minipage}{9cm}
\includegraphics[width=9cm]{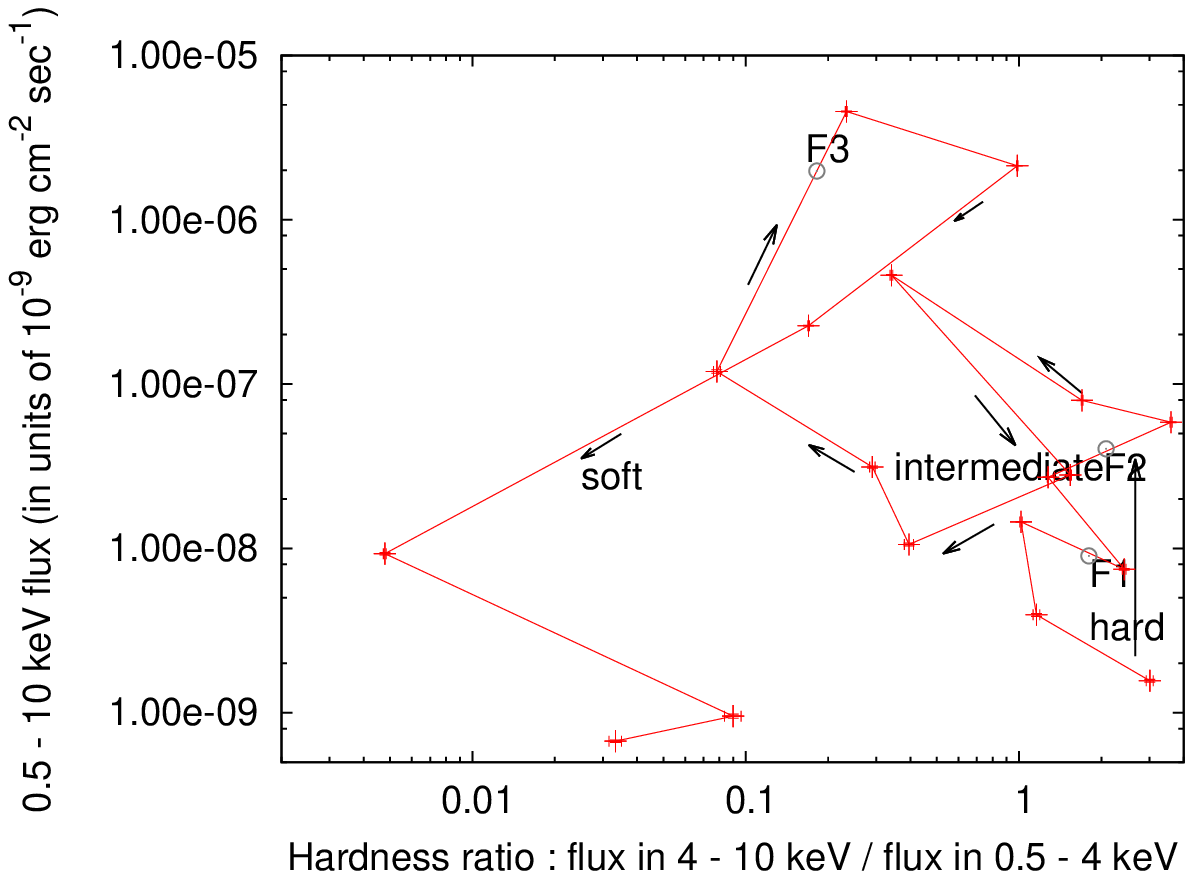}
\end{minipage}
\vskip 0.1in
\begin{minipage}{8cm}
\includegraphics[width=9cm]{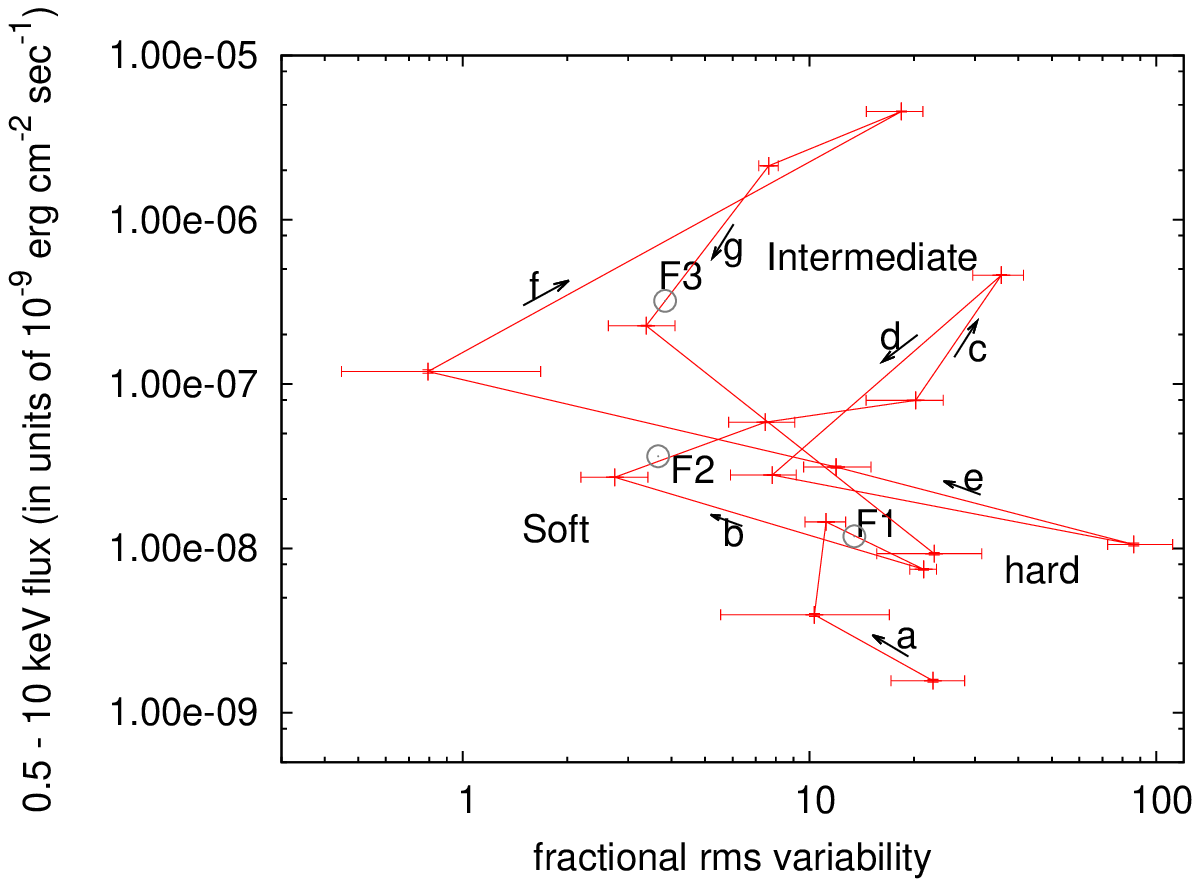}
\end{minipage}
\caption{A plot of the hardness intensity diagram (top panel) and rms-intensity diagram (bottom panel) 
based on XRT data. The hardness ratio has been estimated as ratio of flux in 4 - 10 keV to 0.5 - 4 keV. 
The flux values in different bands have been estimated by excluding the contribution of the 
Fe line emission. The radio flares (F1, F2 and F3) are also indicated as grey circles closer to the 
next X-ray observation. The arrows marked from a to g in the bottom panel indicate the direction of the 
evolution of the RID. The fractional rms variability has been estimated in the frequency 
range of 0.1 - 20 Hz. }
\label{fig5}
\end{figure}

The hardness intensity diagram (HID) and rms-intensity diagram (RID) are generated for XRT observations 
as shown in top panel of Figure \ref{fig5}. Since Fe line emission is observed only for seven 
observations, we have estimated the flux in each energy band by excluding the contribution from the 
Fe line, so as to have uniform estimation 
throughout. We find that during the initial days both the XRT and BAT hardness ratios remain high 
(see, panels d and e in Figure \ref{fig2}) at a lower flux level (top panel of Figure \ref{fig5}). 
Spectral modelling implies the spectra to be dominated by hard emission without any presence of 
a disc emission, and the temporal properties imply that the 
fractional rms variability is higher ($>$ 20\%) during this period (Figure \ref{fig2} and bottom 
panel of Figure \ref{fig5}). The low frequency QPOs observed by INTEGRAL during the 
same period, are similar to the type C QPOs observed in most of the black hole sources 
\citep{Casella2004}. These spectral and temporal characteristics with variations, 
suggest the source to be in the hard state during the initial days.  

The XRT HID is later observed to show random fluctuations with occasional decrease in hardness ratio, 
resulting in a circular track during the intermediate states. For this period, multiple radio 
ejections (marked in grey circles in Figure \ref{fig5}) have occurred and the variations of the 
spectral parameters are random (Figures \ref{fig2} and \ref{fig3}). The fractional 
rms variability is also observed to 
show a random variation (bottom panel of Figure \ref{fig5}) during this period of outburst. 
Although the spectral and temporal parameters suggest that this duration belong to 
an intermediate state, due to the random variation observed it is difficult to differentiate the 
states into hard intermediate and 
soft intermediate. This prohibits the HID to follow the typical `q-shape' 
(\citealt{Belloni2005,2010MNRAS.403...61D} and references therein) observed in other outbursting 
black hole sources like GX 339$-$4 \citep{Belloni2005,Nandi2012}. During the same period, the BAT hardness 
ratio is observed to decrease from 1.25 to 0.3 (panel e of Figure \ref{fig2}).  

As the source flux begins to decline since day 11.99, the hardness ratio is observed to 
decrease (Figure \ref{fig2} and top panel of Figure \ref{fig5}). The XRT HID implies 
that during this period the source transits to soft state, when the fractional rms variability is 
also observed to decrease (bottom panel of 
Figure \ref{fig5}). For the days 12 and 13, although the spectra have hardness ratio 
of $<$0.09, the PDS implies a flat-top noise with higher fractional rms of $>$47\%, and the rms 
variability is observed to be larger up-to a frequency of 60 Hz. Hence we do not indicate 
these observations either in the RID or Figure \ref{fig2}. Thus the XRT HID and RID deviate 
from the typical behaviour, due to the random variation observed during the hard and 
intermediate states\footnote{No random variation has been observed for most of the outbursting 
sources during the initial phase of the outburst, except in SIMS 
(e.g. GX 339$-$4 \citep{Belloni2005,Nandi2012}, XTE J1859$+$226 \citep{HB2005,RN2014}).}. 

\subsection{X-ray characteristics during multiple radio flares}

We observe that during day 3.54, the 0.5 - 10 keV flux increases to 
14.5${\pm 0.07}$ ${\times}$ 10$^{-9}$ erg cm$^{-2}$ sec$^{-1}$ (Figure \ref{fig2}). The XRT 
hardness ratio and fractional rms variability are observed to decrease 
(Figure \ref{fig2}, \ref{fig5}) w.r.t the previous observations. Almost 12 hrs later on day 4 
(i.e. MJD 57192), a radio flare of flux 140 mJy (see inset figure in panel c of 
Figure \ref{fig2}) 
has been observed at 16 GHz using AMI-LA \citep{2015ATel.7658....1M,2015ATel.7714....1M}. We denote this 
flare as F1 in Figure \ref{fig2}, which is rapid and has an oscillating appearance. RATAN-600 also 
reports the detection of a radio flux of 112.25 mJy at 8.2 GHz during the same period 
\citep{2015ATel.7716....1T}. The INTEGRAL observation on day 4.5 has reported an X-ray 
flaring activity (marked I1 in Figure \ref{fig2}) closer 
to this period \citep{2015ATel.7662....1F}. But there 
is no SWIFT observation available on this day. 

We find that the 15 - 150 keV flux, the BAT hardness ratio and XRT fractional rms decreases 
(Figure \ref{fig2}) during day 6.60 with 
respect to the previous observations. It can be also noted from panel c of 
Figure \ref{fig2} that the radio observations indicate an increase in flux to 336.24 mJy (flare F2) at 
8.2 GHz on day 7 \citep{2015ATel.7708....1T,2015ATel.7716....1T}. After the radio flare, 
we find that the 0.5 - 10 keV flux increases to 
460.11${\pm 3.16}$ ${\times}$ 10$^{-9}$ erg cm$^{-2}$ sec$^{-1}$ (panel a of Figure \ref{fig2}) on 
day 7.41, during which both disc temperature and photon index are found to have increased 
(Figure \ref{fig3}). This increase in XRT X-ray flux is in positive correlation with the 
X-ray flare (I2 in Figure \ref{fig2}) observed by INTEGRAL. 

\begin{figure}
\includegraphics[height=9cm,angle=-90]{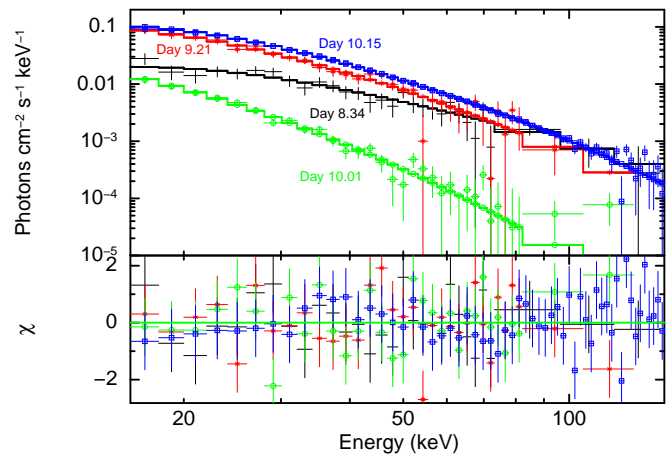}
\caption{Evolution of the 15 - 150 keV spectra during the period of days 8.34 to 10.15, 
indicating decrease in flux almost 24 hrs before the peak radio flare observed on day 11. The 
BAT spectra for days 8.34, 9.21 and 10.01 have been re-binned above 60 keV for better representation.}
\label{fig6}
\end{figure}

Similar characteristics are observed during the peak radio flare (F3 in Figure \ref{fig2}) 
on day 11. In Figure \ref{fig6} we represent the variations of the 15 - 150 keV BAT spectra 
before and after the radio flare. We find that during the observation on day 10.01 (spectra 
indicated in green colour and open circular points in Figure \ref{fig6}), the 15 - 150 keV BAT flux, 
the BAT hardness ratio and the XRT fractional rms variability (panels e and f of Figure \ref{fig2}) 
decreases during this observation. Thus a significant decrease in the hard flux is observed with 
respect to the previous observation on days 9.21 (red star 
points in Figure \ref{fig6}) and 8.34 (spectra in black colour with point type of cross). The radio 
observations indicate an increase in the flux to 4085.1 mJy at 8.2 GHz on day 11, giving rise 
to a peak flare (F3 in Figure \ref{fig2}) with inverted spectrum at low 
frequencies \citep{2015ATel.7667....1T}. Following this event, both XRT and BAT fluxes are 
observed to increase (spectra in blue colour with box type points) and attain a maximum during 
days 10.93 and 11.5 respectively (see Figure \ref{fig2}). INTEGRAL observations also report 
peak X-ray flares (I3 and I4) during this period. The possible connection between the X-ray 
characteristics and radio flares during all these observations are discussed in section \ref{dis}.

\subsection{Evolution of the Fe line emission}

The spectral analysis suggest the presence of an Fe line component during seven observations. In order 
to understand the relativistic effects on this line, we modeled the line profile with \emph{diskline} 
\citep{Fab89} model. The fit results show the parameter of the equivalent width to be of the order of 
10$^{-7}$ keV, which is not appropriate. This also could be giving an indication that the Fe line is not 
originating from the disc and has a different source of origin (see section \ref{dis}). 
In order to consider the broadening of the Fe line, we decide to model it with the 
\emph{gauss} only.
 
\begin{figure}
\includegraphics[height=9cm,angle=-90]{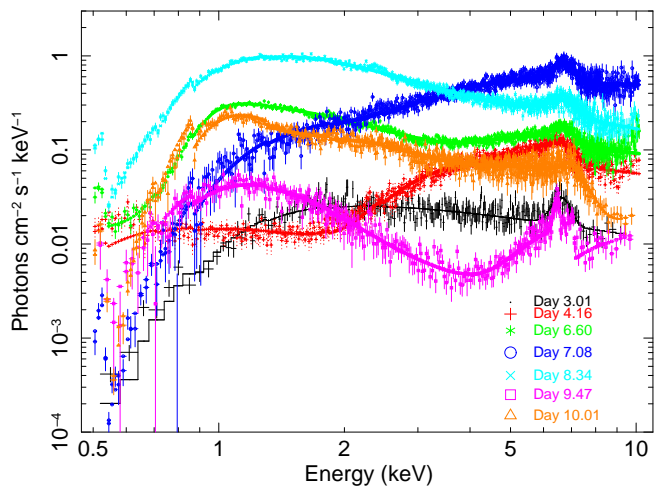}
\caption{Evolution of the Fe line emission in the 0.5 - 10 keV XRT spectra. Although there are twenty 
observations, the signature of Fe line is observed only for seven of these.}
\label{fig7}
\end{figure}

We observe that the peak energy of the Fe line is varying from 6.26 keV to 6.9 keV, and the 
line energy evolves as shown in Figure \ref{fig7}. During 
the first XRT observation on day 3.01, when the source exists in the hard state without any signature 
of the disc component, we observe the Fe line to have an equivalent width of 0.77 keV 
(spectra shown in black colour and dot points in Figure \ref{fig7}). We find that this line contributes 
a fraction of 0.08 of the total flux. The spectral fit for this observation does not show any 
absorption features.  

During days 4.16 to 10.01, the equivalent width of the Fe line is observed to be varying 
between 0.23 keV and 1.15 keV. The fractional contribution of the Fe line flux to the 
total flux, is found to be varying from a minimum of 0.01 to a maximum of 0.12. 
We find that during these observations, the spectral fits also show absorption features, which 
has been taken into account using the \emph{pcfabs} model. This period belongs to the 
intermediate state of the source. The Fe line observed on day 9.47 is found to be very narrow and has 
an equivalent width of 0.23 keV (magenta spectra with box type points in 
Figure \ref{fig7}). We observe from Figure \ref{fig7} that the flux of the Fe line observed on 
days 3.01 and 9.47 are lesser than that for the other observations, with a maximum of 
3.38${\pm0.2}$ $\times$10$^{-9}$ erg cm$^{-2}$ sec$^{-1}$ on day 7.08 (see spectra with blue open 
circles in Figure \ref{fig7}). We discuss the possible reasons for the evolution of the 
Fe line in section \ref{dis}. 

Multiple Fe lines of different energies have been also detected for this outburst of V404 Cyg 
by \citealt{2015ApJ...813L..37K} based on Chandra observations during days 7 and 8. They find 
presence of Fe K${_\alpha}$ of 6.39 keV, K${_\beta}$ of 7.058 keV, Fe XXV of 6.63 and 6.682 keV, and 
Fe XXVI of 6.973 keV. Since XRT has a poor energy resolution and less effective area in comparison 
with Chandra-HETG, we are not able to resolve into the different narrow Fe lines. Yet, it might be 
possible to identify the Fe lines obtained from XRT observations, by comparing their centroid energy 
with the fluorescence lines observed using Chandra. The narrow line which we have observed from the 
XRT observation during day 9.47 at 6.43 keV might be Fe K${_\alpha}$, and that during day 10.01 at 
6.9 keV might be Fe XXVI. The broad lines observed during the other days could be a possible blend 
of some of the other Fe line emissions. 

\section{Discussions and conclusions}
\label{dis}

In this paper, we have studied the spectral and temporal variations of the source 
GS 2023$+$338 (V404 Cyg) based on the SWIFT XRT and BAT observations, during the initial days of 
the June 2015 outburst. It has been understood that outbursting black hole sources undergo 
spectral state transitions during their outburst phase. As the outburst progresses, based on the 
contributions of the thermal (Keplerian disc) and non-thermal emission (Compton corona), 
the source occupies different spectral states of `hard, hard intermediate, soft intermediate 
and soft'. These states can be identified based on the variations of the spectral photon index, 
hardness ratio, the fractional rms variability and the types of QPOs 
\citep{Casella2004, Belloni2005, Deb08, Munoz11, Nandi2012}. During the end of the outburst 
the source passes through the soft intermediate, hard intermediate and hard state, and thus forms 
a `q-profile' in the hardness intensity diagram 
(\citealt{Belloni2005,2010MNRAS.403...61D,Nandi2012,2016A&A...587A..61C} and references therein). 
In this paper, we classify the spectral states of the source V404 Cyg based on the observed variations 
of the spectral and temporal parameters (see section 3.3 and below) with reference 
to \citealt{Belloni2005,2006ARA&A..44...49R,Deb08,Nandi2012}.

These spectral and temporal characteristics, spectral states and evolution of the HID, can 
be understood based on several theoretical models like the two component advective flow model; 
\citep{ST95}. This model consists of a viscous Keplerian flow at the equatorial plane, 
and a less viscous sub-Keplerian flow which sandwiches the Keplerian flow. During the accretion phase, 
the sub-Keplerian flow may undergo a shock transition and form a hot, puffed-up 
dynamical Compton corona which inverse-Comptonize the soft photons from the disc to 
produce hard X-ray photons in the form of a power-law distribution. As the outburst begins, the matter 
in the sub-Keplerian flow dominates giving rise to a hard state with a powerlaw spectrum in the 
higher energies. When the matter flows inward, the flow in the Keplerian disc increases giving 
rise to a multi-coloured disc blackbody emission resulting in the contribution of soft 
photons to increase. This will give rise to the hard intermediate state. As the source reaches the 
soft intermediate state, the contribution of the Keplerian disc emission increases since 
the sub-Keplerian flow decreases. Later while in the soft state the Keplerian emission dominates and 
the sub-Keplerian has minimal contribution to the total flux. Recently, attempts have been
made for several black hole binaries to classify the spectral states based on the instantaneous 
variation of accretion rates of the Keplerian and sub-Keplerian flow, shock parameters etc. 
(see \citealt{2014ApJ...786....4M,2015ApJ...803...59D,2015MNRAS.447.1984D,2016ApJ...819..107J} and 
references therein) within the context of this two component model. Similar studies for V404 Cyg using this model will be performed later and the results will be presented elsewhere.

Based on these general understandings of outbursting black hole binaries, we study the 
properties of the source V404 Cyg during its June 2015 outburst. Observations by INTEGRAL 
reported by \citealt{2015A&A...581L...9R} suggests the presence of a cut-off powerlaw 
describing the thermal Comptonization process and additional powerlaw at energies above 100 keV. They 
associate the flaring activity with the hard X-ray spectral changes and note that the source remains 
harder during the X-ray flares. In this paper, we find that the variations observed for V404 Cyg in 
the spectral parameters, X-ray flux, hardness ratio and the fractional rms variability 
suggest that the source has occupied different spectral states during the outburst.   

We observe that during the initial days of the outburst, the spectral 
characteristics suggest a dominant hard emission (Figure \ref{fig2}), described by a powerlaw without 
any signature of cut-off and reflection components 
in the spectra. The hardness ratio is observed to remain high and the 
HID indicates a random variation during these observations. The temporal characteristics also imply 
that the fractional rms variability remains high during these days (Figures \ref{fig2},\ref{fig5}). 
Thus the spectral and temporal properties during this period suggest the source to have 
occupied a hard state. 

The INTEGRAL satellite which has larger effective area and better 
time resolution has observed significant QPOs on days 0 and 2 \citep{2015ATel.7726....1P} during which 
there are no SWIFT-XRT observations. The PDS of BAT observation on day 0 also does not show presence 
of QPOs. Till date there has been no detailed description of the INTEGRAL observations of QPOs in 
a manuscript, except an Astronomer's Telegram \citep{2015ATel.7726....1P} which we have 
referred in this manuscript. Contrary to \citealt{2015ATel.7694....1M}, we do not observe a 
significant QPO at 1.7 Hz during day 3.03 (i.e. MJD 57191.03) of the SWIFT observations but a weak 
peaked component is noted at lower frequencies (see Figure \ref{fig4} and section 3.2).

Observations after 3 days imply a significant contribution to the spectrum from the disc 
(see bottom panel of Figure \ref{fig3}, Table \ref{all}), along-with 
the hard emission which exhibits a cut-off and reflection component between 15 keV and 30 keV. 
This presence of a soft disc/spectral softening is not observed by INTEGRAL/FERMI probably 
due to their energy range coverage. We observe intermittent absence of both the disc and 
the Fe-line emission during the outburst. The hardness ratio is observed to 
decrease occasionally during this period (panels d and e of Figure \ref{fig2}). The hardness intensity 
diagram, shows random fluctuations (top panel of Figure \ref{fig5}) and hence do not exactly follow 
the `q-shape' observed for most of the black hole transients 
(\citealt{Belloni2005,2010MNRAS.403...61D,Nandi2012,RN2014} and references therein). 

The fractional rms variability is observed to exhibit a random variation as shown in bottom 
panel of Figure \ref{fig5} which is dissimilar to that for most of the other sources 
like GX 339$-$4 \citep{Munoz11}. Unlike the 1989 
outburst where the power density spectra were observed to show a flat-top noise 
\citep{1997A&A...321..776O}, for the 2015 outburst we find that the PDS have either 
broad-band noise or flat-top noise for most of the observations. The power spectrum is 
observed to exhibit powerlaw nature for a few observations 
(i.e. during days 3.54, 7.08 and 11.52), and also a decrease in fractional rms variability. 
All these characteristics suggest the period of days 3.54 to 11.9, to belong to an intermediate 
state. Weak mHz QPOs are detected based on FERMI-GBM observations during days 4 
and 5 \citep{2016arXiv160100911J} while in this state. We find that during the near 
simultaneous (w.r.t the FERMI observations) XRT observation on day 4.15, there is a weak peaked 
component present (see third panel of Figure \ref{fig4}).

The source is observed to attain its peak during this intermediate state. Based on BAT observations, 
it has been reported by \citealt{2015ATel.7755....1S} that the maximum 
luminosity of the source is of the order of 1.6${\times}$10$^{39}$ erg sec$^{-1}$ in 1 - 500 keV. 
We also obtain a value of 1.4${\times}$10$^{39}$ erg sec$^{-1}$ in the same energy band 
using \emph{dummyrsp} in XSpec for the same BAT observation used by \citealt{2015ATel.7755....1S}. 
Yet, when we incorporate the XRT spectrum, a significant contribution from 
the Keplerian disc (i.e. soft emission) is also observed to exist. Hence the total 
luminosity observed should be estimated by considering both the soft and hard emission spectrum, 
and it is found to be 1.68${\pm 0.08}$ ${\times}$10$^{39}$ erg sec$^{-1}$ (see section 
\ref{res} in 0.5 - 150 keV energy band. This value of the peak 
luminosity is observed to match with the Eddington luminosity expected 
for a source with mass $>$ 10 M$_{\odot}$. \citealt{1994MNRAS.271L..10S} has reported the 
source mass to be of 12 M$_{\odot}$. 

Observations since day 12 suggest that the hardness ratio decreases to its minimum 
(Figures \ref{fig2} and \ref{fig5}), and the contribution from hard emission also 
declines (see Figure \ref{fig2} and Table \ref{all}). Hence this period probably belongs to the 
soft state. We observe that during this outburst of the source, multiple radio flares have also been 
detected, as shown in Figures \ref{fig2}, \ref{fig3} and \ref{fig5}. It is found that although on 
days 12 and 13 the spectra is softer, the PDS imply higher rms variability. This huge 
increase in rms power might be due to the dust scattering rings observed during this 
period \citep{2015ATel.7736....1B}.

We find that just before the detection of a radio flare, the spectral and temporal properties 
are changing. The energy spectra is observed to become softer indicated by either an 
increase in soft/XRT (0.5 - 10 keV) flux or decrease in hard/BAT (15 - 150 keV) flux, and the 
temporal properties imply a reduction in the fractional rms variability 
(Figures \ref{fig2} and \ref{fig5}). This has been observed three times
for this particular outburst of the source V404 Cyg. All the flares F1, F2 and F3 are preceded by such 
variations, and the X-ray flare is observed to attain a maximum following the 
event. \citealt{2015ATel.7658....1M,2015ATel.7714....1M,2015ATel.7708....1T,2015ATel.7716....1T} 
have suggested/reported these flares to be optically 
thin\footnote{http://www.sao.ru/Doc\_en/SciNews/2016/Trushkin} at higher frequencies and observed 
them to be similar to the bright ejections from GRS 1915$+$105. Such changes in X-ray characteristics 
have been observed earlier for several Galactic black hole transients like XTE J1859$+$226, 
XTE J1550$-$564, H 1743$-$322 (see \citealt{FBG04,FHB09,MJ2012,RN2014,RNVS16} and 
references therein).  

It has been understood for many black hole sources that the jet ejection/radio flare might be 
occurring due to the ejection of matter from the dynamical Compton 
corona (\citealt{1999A&A...351..185C,SV2001,Nandi2001} and references therein). This will result 
in the decrease of hard photons and fractional rms variability, and QPOs may not be observed 
in the power spectra. Detailed studies on this have been performed for GRS 1915$+$105 
by \citealt{SV2001}, and for other outbursting sources like XTE J1859$+$226, H 1743$-$322 etc. 
\citep{FBG04,FHB09,MJ2012,RN2014,RNVS16,Nandi16}.

The disc-jet coupling phenomenon has been understood based on theoretical studies 
by \citealt{BZ1977} and using simulations by \citealt{MG2004, dV2005} considering the black hole spin 
as the origin for jets, or based on poloidal magnetic field by \citealt{VarTag}. Further 
understanding of the jet ejection as mass loss from a hydrodynamical accretion disc due to 
the presence of centrifugal barrier \citep{1999A&A...351..185C}, and the collimation of jet 
\citep{Spru96,Fuk01,Chat04,Chat05} have also been studied in detail. \citealt{FBG04} have 
developed an unified model of the disc-jet coupling based on the observational results. 
The model states that a compact non-relativistic jet exists while in the hard and hard intermediate 
states, and during the transition from hard intermediate to soft intermediate state, 
a relativistic jet ejection occurs (see section \ref{intro} also).  

Multiple radio flaring/jets have been observed during the transition to, and while in the soft 
intermediate states in other black hole transients like XTE J1859$+$226 \citep{FHB09,RN2014}, 
XTE J1752$-$223 \citep{Brock13}. But for 
V404 Cyg, we observe multiple radio flaring just after the transition from the hard state to the 
intermediate state. The first radio flare (F1) which appears to be oscillating/variable, probably 
occurs during the transition from the hard state. Before this radio flare both the soft
(XRT) and hard (BAT) flux increases (panels a and b of Figure \ref{fig2}),
and this is because the dynamical Compton corona is relatively large in size and hot, 
so that even in the presence of outflow/jet (which is not very strong in this case), the disc 
is capable of producing hard energy and complete evacuation of the dynamical corona 
is not possible. Several works have been performed based on simulations, in order to 
understand the characteristics of variable/oscillating outflows/flares similar to the flare 
F1 of V404 Cyg (see \citealt{1996ApJ...470..460M,2001ApJ...563L..57M,dcnm14} and references therein).

The second and third flares (F2 and F3 respectively) are much stronger than the first flare (F1) 
and happen while in the intermediate states. In this case the disc is much closer 
(smaller corona) to the central engine, and matter will be evacuated from the 
dynamical corona in the form of jet/outflow. 
Hence both the soft and hard X-ray contribution decreases and the HID indicates X-ray 
radiation becomes relatively softer (see top panel of Figure \ref{fig5}, and Figure \ref{fig6}). Also we 
see X-ray flaring activity after the radio flare which can be understood in a two-component 
accretion flow paradigm (\citealt{ST95}, discussed earlier). Here, the sub-Keplerian 
flow has a much larger radial velocity than the Keplerian flow. Hence, after the radio 
flaring in intermediate state, quickly the evacuated dynamical corona is going to be 
filled by hot sub-Keplerian flow and this increases the hard X-ray activity. But after the 
first radio flare, we do not see any X-ray flare because the disc is more steady with 
higher rate of sub-Keplerian matter.

The random variation occurring in the HID could be probably due to some change of the 
accretion dynamics resulting in the ejection of multiple radio flares. It might be also happening 
that the hybrid flow comprising of Keplerian and sub-Keplerian \citep{ST95} are being 
changed drastically due to the sudden change at the outer edge of the disc driven by some 
peculiar behaviour of the binary companion (see also \citealt{2015ARep...59..447C} and 
references therein, for possible effects of the companion). To have a better understanding of 
the spectral and temporal behaviour of this source, 
one needs to do a more realistic modeling, which is beyond the scope of this paper, 
and calculate the flow parameters (say accretion rates, size of the central region, 
temperature of corona etc.; for an 
example see \citealt{2015MNRAS.447.1984D,2015ApJ...807..108I,2016MNRAS.tmp..664M} and 
references therein for a few other sources). This may explain the random behaviour of 
HID in hard and intermediate states for this source. In fact, we do see some randomness in 
HID for outbursting X-ray binaries in intermediate states, because both Keplerian and 
sub-Keplerian accretion rates are comparable 
(see \citealt{2010ApJ...710L.147M,2015MNRAS.447.1984D,2016MNRAS.tmp..664M} and references therein)  
or in other sense supply of hot electrons and soft photons are comparable which introduces non-linearity 
into the system. Our recent study \citep{RNVS16} of sources like GX 339$-$4, XTE J1859$+$226, 
H 1743$-$322 suggests the random variation during the transition from hard intermediate to 
soft intermediate state to be a `local' phenomenon which results in ejection of radio flares. 
In V404 Cyg, we understand that the random variations are occurring for the hard and intermediate 
states also, and probably the change in accretion dynamics is taking place globally w.r.t the disc 
system, along-with `local' changes which causes the ejection of matter in the form of radio flares.

It is observed that a few moments after the huge radio ejection (F3), the X-ray spectra becomes 
harder and the disc component is not observed. The spectral hardening might be occurring due 
to re-filling of matter in the dynamical corona. It is also possible that during the jet 
ejection, a portion of the inner part of the Keplerian disc has been ejected out resulting 
in non-detection of the disc component in the subsequent observation. We observe 
the disc emission again after a few hours during the next observation, since the 
Keplerian disc takes viscous timescale to refill and hence 
the re-emergence of soft/thermal emission. So more co-ordinated and dedicated multi-wavelength 
observations during massive radio flares in black hole sources can help to have a better 
understanding of similar observational features.

We find that during the observations when an Fe line is present while in the intermediate state, 
there are signatures of absorption for the low energy XRT spectra indicated by a partial covering 
absorption. This suggests the presence of an outflow/wind along-with the disc and corona. 
The Fe line observed based on our SWIFT analysis is found to have its peak energy varying 
from 6.26 keV to 6.9 keV. Presence of multiple Fe line emissions during the intermediate 
state of this source has been observed by Chandra-HETG \citep{2015ApJ...813L..37K} due to 
its very high spectral resolution in comparison to the SWIFT XRT. This variation of the Fe line 
energy suggests that the wind emission is strong enough to ionize the matter, and result in 
the emission of Fe lines at different energies. 

Although the Fe-line obtained from our XRT analysis do not have a constant energy during the outburst 
but are varying randomly, we find that these lines differ from each other in their equivalent width 
and the flux. Those Fe lines which are observed after a radio flare, have more flux than the other 
lines. For example, the Fe line observed on day 4.16 has 10 times more flux than on day 3.01, and 
the observation of day 4.16 occurs after the first radio flare F1. Similarly the flux of Fe line on 
day 7.08 which takes place after the radio flare F2, is observed to be more than the 
previous observation. This suggests that stronger Fe line emission is associated with the wind
emission, probably not related with the jet activity. 
We also note that the \emph{diskline} model could not give a correct estimate of the line width, and 
this might have happened because the origin of the Fe line is not related to the disc.  

During day 3.01 when the Fe line is observed in the hard state (where there is no detection 
of disc emission), probably the wind is optically thin 
and weak, that the spectra does not indicate strong signatures of absorption. 
\citealt{1999A&A...351..185C,2014Ap&SS.353..223M} suggest that the wind outflows can be originated 
from the centrifugal barrier of the two component model during the hard state of a source, 
and the rate of outflow 
is positively correlated with the accretion rate. Hence the Fe line observed during the hard state 
of V404 Cyg suggests its origin to be related to the wind outflow occurring from the centrifugal 
barrier. Possible mechanism for generation of winds due to heating of shock waves at the post shock region (centrifugal barrier) has been previously discussed in detail by \citealt{1994ApJ...425..161M} and references therein. Thus we understand that it is possible that wind/outflow is responsible for the Fe line 
emission observed during both the hard and intermediate states and, it may not be due to fluorescence 
similar to that observed for most of the outbursting black hole sources. 
\citealt{ST95,2003ApJ...598..411T} have discussed in detail about the possible connection between 
reflection profiles of Fe line and origin of wind outflows in stellar mass black hole binaries.

Most of the black hole binaries have an orbital period of a few hours, except a 
very few which have the orbital period in days \citep{2006ARA&A..44...49R}. The sources 
like XTE J1118$+$108, A 0620$-$00, XTE J1550$-$564 have orbital period in the range of 4 hrs to 37 hrs 
with their accretion rate varying from 10$^{15}$ g sec$^{-1}$ to 10$^{16}$ g sec$^{-1}$. But 
sources which do not have an exact `q-shape' in their HID have longer orbital periods. For an example, 
GRS 1915$+$105 has an orbital period of 33.5 days with peak accretion rate 
of 3.4${\times}$10$^{18}$ g sec$^{-1}$. Although the source IGR J17091$-$3624 is similar to 
GRS 1915$+$105 in the variabilities it exhibits, the peak accretion rate is found to be lesser than 
that of the latter \citep{2015ApJ...807..108I}. We understand that V404 Cyg which has an orbital 
period of 6.47 days \citep{1992Natur.355..614C}, is probably not similar to most of the black 
hole sources (due to the deviation from `q-shape' in HID and other characteristics discussed above) 
having a peak accretion rate of 3$\times$10$^{18}$ g sec$^{-1}$ during its 2015 outburst. These 
findings show that, in low mass X-ray binaries, the sources which have shorter orbital 
period are having lower accretion rate than the sources which have longer orbital period 
and high accretion rate. But for the source Cygnus X-1 which belongs to a high mass 
X-ray binary with a longer orbital period of 5.9 days \citep{1999A&A...343..861B}, the peak 
accretion rate is only 6.7${\times}$10$^{16}$ g sec$^{-1}$. This suggests that probably the accretion 
rate of a source depends not just on the orbital period but also on the characteristics of the 
binary system. 

Based on the optical observations \citealt{2016Natur.529...54K} have observed that during the 2015 
outburst of the source V404 Cyg, due to its long orbital period, the accreted matter achieves the 
critical density for thermal disc instability at a smaller radius of the disc. The black hole sources
 with shorter orbital period of a few hours will have the 
thermal disc instability occurring at a larger outer radius of the disc. 
Rapid optical variations giving rise to oscillations are observed in the optical lightcurve of 
V404 Cyg by \citealt{2016Natur.529...54K} which probably is due to the longer orbital period of 
the source. Although similar oscillations have been observed in the X-ray lightcurves of 
GRS 1915$+$105 and IGR J17091$-$3624, the reason/origin for the X-ray and optical oscillations might 
be different. It might be possible that the phenomenon responsible for the origin of the optical oscillations gets perturbed during the accretion process and is reflected in X-rays.
    
The random variation of the spectral parameters, hardness ratio, fractional rms 
variability, multiple radio flares, absorption features and the fact that the HID does not 
follow a `q-profile' implies that the source V404 Cyg has a complex characteristic. These 
deviations categorize this source different from most of the outbursting 
black hole sources. 

Thus, based on the spectral and temporal properties of the source GS 2023$+$338 (V404 Cyg) 
during its 2015 outburst, we arrive at the following conclusions. 

\begin{itemize}
\item The spectral and temporal characteristics and the HID implies that the source occupies 
hard, intermediate and soft spectral states during this outburst.

\item The HID deviates from the typical q-shape observed in most of the black hole transients. 
We find that the HID and the RID, have a random pattern while in the hard and intermediate 
states, and this also belongs to the period of multiple radio and X-ray flares. These observations 
imply that the source has a complex characteristic, unlike other black hole sources 
like GX 339$-$4, H 1743$-$322 etc.. 

\item We observe weak peaked components in the power density spectra during the intermediate state of the source. 

\item We find that just before the detection of a radio flare/jet ejection, the energy spectra softens 
and the fractional rms variability decreases. This correlation between X-ray and Radio characteristics indicates the ejection of matter from the corona into jets. 

\item Absence of a disc is indicated just after the peak radio flare. This might be due to the possible evacuation of the inner part of the Keplerian disc. 

\item The spectral evolution indicates presence of absorption features for the low energy XRT spectral 
continuum, probably occurring due to outflow/wind emission. The evolution of the variable Fe line 
observed using SWIFT, during both the hard state (which has no signature of Keplerian disc 
emission) and the intermediate state, suggests that probably its origin is related to the wind, 
and not to the fluorescence emission from disc.

\end{itemize}
All these properties, imply that the source has a complex evolution during its 2015 outburst.

\section*{Acknowledgements}
This research has made use of the data obtained through High Energy Astrophysics Science Archive Research 
Center on-line service, provided by NASA/GSFC. We are thankful to A. Beardmore for useful suggestions 
related to SWIFT-XRT data reduction. AN thanks GD, SAG, ISAC for continuous support to carry out this 
research. We thank the reviewer whose comments and suggestions have helped to improve the quality of 
this manuscript.

\bibliographystyle{mnras} 
\bibliography{biblio}

\appendix
\begin{table*}
\caption{Log of SWIFT XRT and BAT observations considered for analysis}
\label{tab-appendix}
\begin{tabular}{cccccc}
\hline\\
Observation ID & Date & Time & MJD & \multicolumn{2}{c}{Exposure time (in sec)} \\
     &     &       & & XRT & BAT \\
\hline\\

00643949000&2015-06-15& 18:35:08&	57188.77&  - & 1202.12 \\
00031403038&2015-06-18& 00:23:00&	57191.01& 610.5 & - \\
00644520000&2015-06-18& 00:45:29&	57191.03& 216.7 & 1202.05\\
00644627000&2015-06-18& 13:10:23&	57191.54& 2303 & 1202.06\\
00031403042&2015-06-19& 03:48:50&	57192.15& 1262&-\\ 
00645176000&2015-06-20& 13:24:38&	57193.55& - & 542.11 \\
00031403046&2015-06-21& 14:30:58&	57194.60& 240 & 6.85 \\
00031403045&2015-06-22& 02:08:12&	57195.08& 930.3 & -\\
00031403049&2015-06-22& 08:32:24&	57195.35& 2976.6 & -\\
00031403047&2015-06-22& 09:59:31&	57195.41& 1902.9 & -\\
00033832001&2015-06-23& 08:04:12&	57196.33& 1867.3 & 8.16 \\
00031403052&2015-06-23& 21:27:34&	57196.89& 275.05 & -\\
00031403054&2015-06-24& 05:01:51&	57197.20& - & 4.6 \\
00033832002&2015-06-24& 11:16:25&	57197.46& 1519.5 & -\\
00031403055&2015-06-25& 00:05:45&	57198.00& 1028.5 & -\\
00031403056&2015-06-25& 00:23:15&	57198.01& 818.5 & 713.2 \\
00031403057&2015-06-25& 03:41:26&	57198.15& - & 745.6 \\
00031403058&2015-06-25& 22:17:00&	57198.92& 1312.4 & -\\
00033861001&2015-06-26& 12:37:36&	57199.52& 1482.7 & 703.4 \\
00031403060&2015-06-26& 23:49:54&	57199.99& 1417.9 & 1233.0 \\
00033861002&2015-06-27& 09:23:56&	57200.39& 1472.1 & -\\
00031403062&2015-06-27& 22:10:25&	57200.92& 1574.4 &-\\
00031403064&2015-06-28& 23:43:58&	57201.99& 1552.1 &-\\
\hline
\end{tabular}
\end{table*}

\bsp	
\label{lastpage}
\end{document}